\RequirePackage{fix-cm}
\documentclass[smallextended]{svjour3}       
\smartqed  

\usepackage{graphicx,epsfig}
\usepackage{upgreek}
\usepackage[version=3]{mhchem}
\usepackage{lmodern} 
\usepackage[T1]{fontenc} 
\usepackage{multirow}
\usepackage{tabularx}

\journalname{Transport in Porous Media}

\newcommand{\dC}{\,$^{\circ}$C}

\newcommand{\degr}{$^{\circ}$}

\newcolumntype{C}[1]{>{\centering\arraybackslash}p{#1}}

\usepackage{color}
\newcommand{\rem}[1]{\textcolor{blue}{}}

\begin{document}

\title{Capillarity-Driven Oil Flow in Nanopores: Darcy Scale Analysis of Lucas-Washburn Imbibition Dynamics}

%


\titlerunning{Capillarity-Driven Oil Flow in Nanopores}        

\author{Simon Gruener         \and
        Patrick Huber 
}


\institute{Simon Gruener \at
              Sorption and Permeation Laboratory, BASF SE, D-67056 Ludwigshafen (Germany) \\
              \email{simon-alexander.gruener@basf.com}           
           \and
           Patrick Huber \at
              Institute of Materials Physics and Technology, Hamburg University of Technology, D-21073 Hamburg-Harburg (Germany)
               \email{patrick.huber@tuhh.de}  
}

\date{Received: date / Accepted: date}

\maketitle

\begin{abstract}
We present gravimetrical and optical imaging experiments on the capillarity-driven imbibition of silicone oils in monolithic silica glasses traversed by 3D networks of pores  (mesoporous Vycor glass with 6.5 nm or 10 nm pore diameters). As evidenced by a robust square-root-of-time Lucas-Washburn (L-W) filling kinetics, the capillary rise is governed by a balance of capillarity and viscous drag forces in the absence of inertia and gravitational effects over the entire experimental times studied, ranging from a few seconds up to 10 days. A video on the infiltration process corroborates a collective pore filling as well as pronounced imbibition front broadening resulting from the capillarity and permeability disorder, typical of Vycor glasses. The transport process is analyzed within a Darcy scale description, considering a generalized pre-factor of the L-W law, termed Lucas-Washburn-Darcy imbibition ability. It assumes a Hagen-Poiseuille velocity profile in the pores and depends on the porosity, the mean pore diameter, the tortuosity and the velocity slip length and thus on the effective hydraulic pore diameter. For both matrices a reduced imbibition speed and thus reduced imbibition ability, compared to the one assuming the nominal pore diameter, bulk fluidity and bulk capillarity, can be quantitatively traced to an immobile, pore-wall adsorbed boundary layer of 1.4~nm thickness. Presumably, it consists of a monolayer of water molecules adsorbed on the hydrophilic pore walls covered by a monolayer of flat-laying silicone oil molecules. Our study highlights the importance of immobile nanoscopic boundary layers on the flow in tight oil reservoirs as well as the validity of the Darcy scale description for transport in mesoporous media. 

\end{abstract}

\section{Introduction}

\begin{figure} \center
\includegraphics[angle=90,width=1\columnwidth]{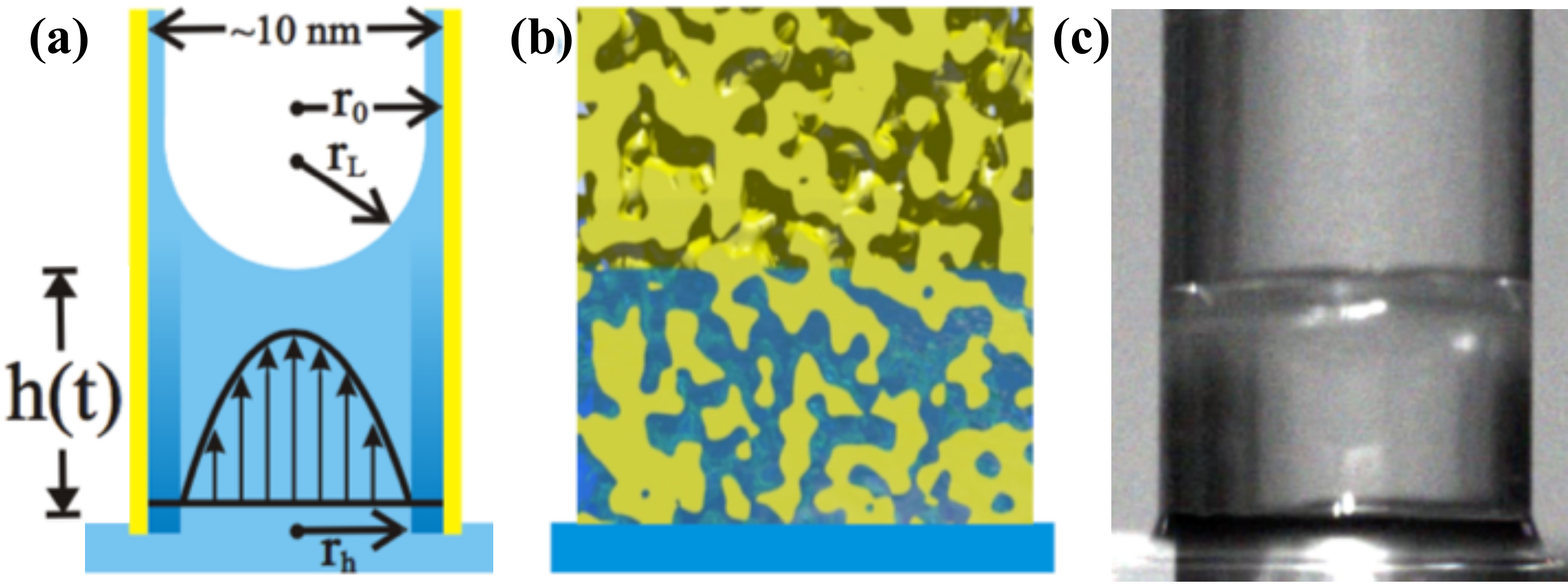}
\caption{Capillary Rise of Oil in Nanopores. (a) Illustration of capillarity-driven liquid imbibition in a cylindrical capillary with a preadsorbed water layer. The fluid has advanced up to the height $h(t)$ and a parabolic\rem{fluid} velocity profile\rem{along with preadsorbed water layers beyond $h(t)$ and} {developed while} a\rem{shaded resting} boundary layer {(shaded region) remains at rest. The different radii are discussed in the text. (b) Raytracing side-view on spontaneous imbibition in a Vycor monolith, which is represented by a clipped Gaussian random-field. (c) Picture of a V5 Vycor rod in which a silicone oil imbibition front has advanced up to a height $h(t)= 5$~mm. See also a video on the imbibition process in the supplementary.} 
}
\label{fig:VycorImbIllust}
\end{figure}

Fluid transport in pores a few nanometers across is of relevance in many natural and technological processes \cite{Eijkel2005}\cite{Stone2004}\cite{Squires2005}\cite{Schoch2008}\cite{Kirby2010}, ranging from water transport in soils \cite{Alyafei2016}\cite{Bao2017}, plants \cite{Stroock2014}\cite{Zhou2018} and biomembranes \cite{Zhang2018} to water filtration, catalysis \cite{Montemore2017}, print\cite{Kuijpers2018} and Lab-on-a-Chip \cite{Piruska2010}\cite{Li2017} technologies. It is also of increasing importance in the synthesis of hybrid materials \cite{Sousa2014}, \cite{Martin2014} \cite{Busch2017} by melt-infiltration \cite{Jongh2013}. In particular, self-propelled, capillarity-driven imbibition in nanoporous media plays a dominant role in many petrophysical processes, ranging from the mass transfer in fractured reservoirs during a waterflood to wettability characterization of rock samples and geothermal reservoirs. \cite{Schmid2013} \cite{Huber2015}\cite{Meng2017} \cite{Yang2017}\cite{Zhang2018a}

The advent of tailorable nanoporous materials, most prominently based on carbon \cite{Holt2006}, silicon  \cite{Vincent2017}, \cite{Gruener2008}, gold \cite{Xue2014}, silica with sponge-like \cite{Gruener2009}\cite{Kelly2018}\cite{Kiepsch2016}, \cite{Gruener2012} and regular pore geometry\cite{Persson2007}\cite{Sentker2018}, and alumina \cite{Shin2007} \cite{Koklu2017} \cite{Yao2017} provides model porous media in order to study this phenomenology in well-defined spatial confinement \cite{Kriel2014}. 

Because of the extreme spatial restrictions in nanopores the validity of continuum hydrodynamics is questionable, both with regard to the coarse-graining procedure as well as the correctness of the standard no-slip velocity boundary condition at the pore wall.\cite{Falk2015} The details of the velocity profile in the proximity of the confining walls sensitively determine the overall transport rates. The ``no-slip at the wall'' concept is considered to hold for a single-component fluid, a wetted surface, and low levels of shear stress.\cite{Bocquet2014} In many engineering applications these conditions are not met.\cite{Stone2004} Both experimental and theoretical studies have revealed that slippage, that is a finite velocity of the liquid at the wall can occur in systems with surfactants, at high shear rates, low roughnesses of the confining walls as well as crystalline wall structures incommensurable with adsorbed monolayers of the respective liquid. \cite{Thompson1997}\cite{Cieplak2000}\cite{Schmatko2005}\cite{Neto2005}\cite{Servantie2008}\cite{Sendner2009}\cite{Baeumchen2012}\cite{Ortiz-Young2013}\cite{Bocquet2014}\cite{Gruener2016}\cite{Secchi2016}. 

Moreover, liquids slipping at a substrate are observed in non-wetting configurations\cite{Vinogradova99}\cite{Lasne08}\cite{Barrat99}\cite{Baeumchen2012}\cite{Gruener2016b}\cite{Wu2017}\cite{Meng2017}. Also applying chain-like or more generally spoken polymeric fluids seems to facilitate the occurrence of slip at the fluid-solid interface \cite{Binder07}\cite{Gruener2016}.

The first experiments to explore flow behaviour through mesoporous glasses were performed by Nordberg \cite{Nordberg1944}, and Debye and Cleland in the mid of the last century \cite{Debye1959}. For liquid hydrocarbons flow rates in agreement with the classical Hagen-Poiseuille prediction for simple capillaries were observed, if an adsorbed layer of molecular thickness at the wall is considered in the transport process. By contrast, for even smaller pores, below 2~nm, as typical for kerogen transport in shales, a break-down of classical hydrodynamic concepts, in particular the Darcy scale description as a generalization of the Hagen-Poiseuille law towards viscous flows in complex pore networks is expected \cite{Falk2015}.  

Depending on the relative size of the pores and the basic building blocks of the liquid, the temperature and the pressure of the fluid, transport in porous media can be governed by a complex interplay of adsorption processes, Knudsen, \cite{Gruener2008}\cite{Kiepsch2016}\cite{Wang2016}, Fickian and surface diffusion \cite{Yamashita2015}, as well as viscous liquid flow driven by capillarity (``spontaneous imbibition'') or by hydraulic pressure (``forced imbibition'') \cite{Gruener2016b}\cite{Vincent2017}. 

We focus here on spontaneous imbibition in porous silica monoliths with pore diameters of 6.5 and 10~nm and thus in the lower ''mesoporous'' regime. Previous studies on the flow of water \cite{Huber2007} \cite{Gruener2009a}\cite{Gruener2016b}, and of linear hydrocarbons \cite{Gruener2009b}\cite{Gruener2016} in such porous glasses revealed a retained fluidity and capillarity compared to the bulk liquids, if a sticky molecular boundary layer is assumed at the pore walls. It results in a slow-down of the imbibition dynamics. In this study, we extend our previous spontaneous imbibition studies on simple liquids towards slightly more complex molecules, i.e. silicone oils. We perform gravimetrical experiments and analyze the scaling of the measured imbibition kinetics based on a Darcy effective medium ansatz.

\section{Materials and Methods}

\subsection{Porous Glass Substrates}
Nanoporous glass samples were purchased from Corning glass (Vycor glass, code 7930). Vycor is virtually pure fused silica glass permeated by a three-dimensional network of interconnected pores \cite{Levitz91}\cite{Huber1999}. It is formed by a leaching process after spinodal decomposition of a borosilicate glass. Therefore, its geometric structure can be well represented by Gaussian random-fields \cite{Gommes2018}, see Fig. \ref{fig:VycorImbIllust} for a raytracing illustration of the Vycor structure.  
The experiments were performed with two types of Vycor with identical porosity $\phi_0 \approx 0.3$ and differing mean pore radius $\overline{r}_0$. The two types will be termed V5 ($\overline{r}_0=3.4$~nm, $\phi_0$=0.315$\pm$0.005) and V10 ($\overline{r}_0=5.0$~nm, $\phi_0$=0.3$\pm$0.005) in the following. The characterization of the matrix properties, in particular $\overline{r}_0$, relies on  volumetric nitrogen sorption isotherms performed at 77~K.\cite{Gruener2016} 

The relatively low porosity along with the high elastic modulus of the silica glass renders any liquid-uptake-induced mechanical deformation negligible \cite{Gor2017}, in the sub-percent range. Any impact of swelling or contraction, which in principle could affect the imbibition process significantly \cite{Kvick2017} can thus be ignored. 

We cut regularly shaped cylinders and blocks of height $d$ ($\sim 10$~mm) from the delivered rods. To remove any organic contamination we subjected them to a cleaning procedure with hydrogen peroxide and nitric acid followed by rinsing in deionized Millipore water and drying at 200~\degr C in vacuum for two days. The pore walls of Vycor are polar due to a silanol termination. This hydroxylation renders these matrizes highly hydrophilic.\cite{Gruener2009}

\subsection{Silicone Oils}

\begin{figure}[!t]
\centering
\includegraphics*[width=.5 \linewidth]{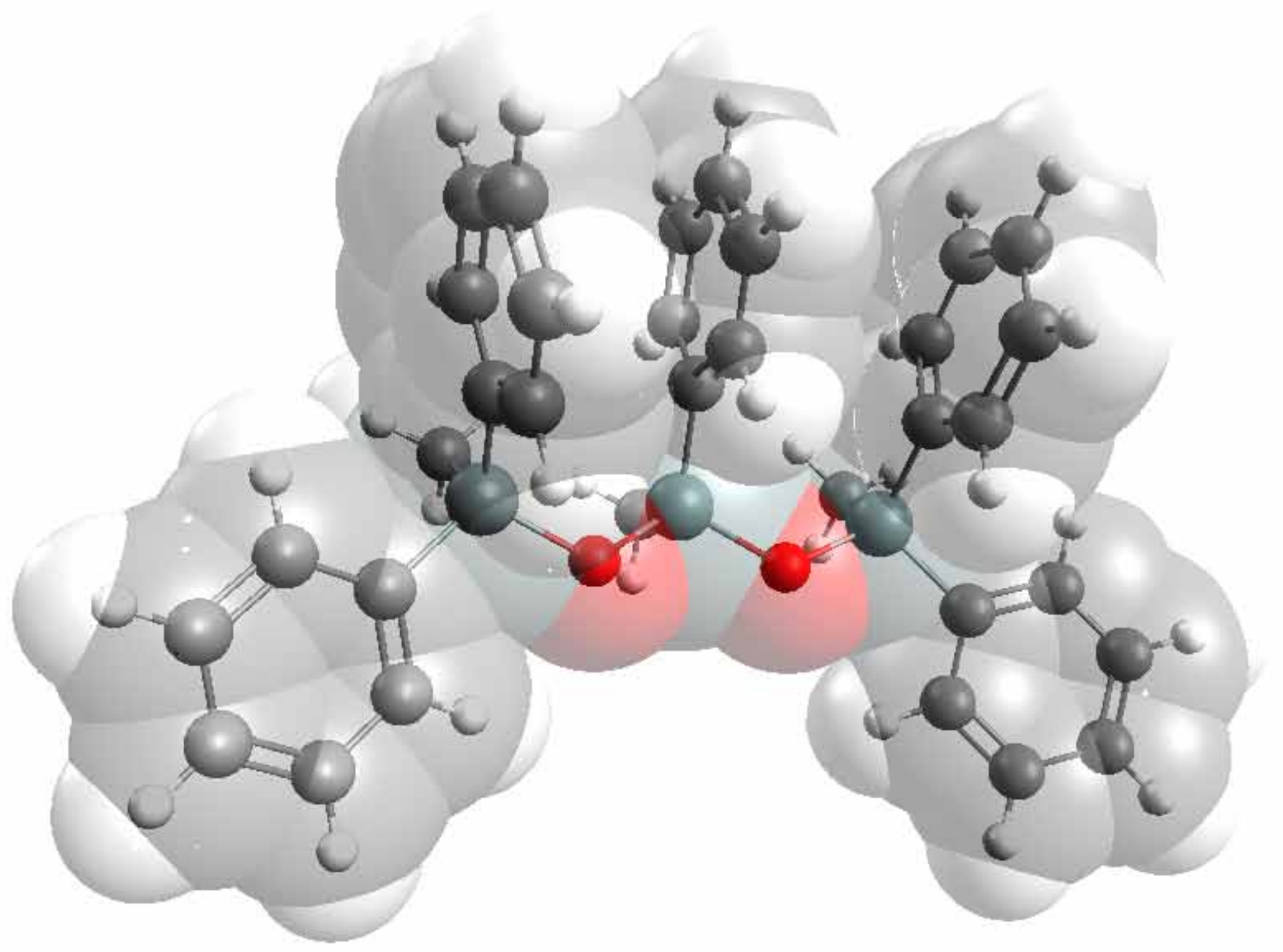}
\caption[]{Molecular structure of silicone oil. Illustration of the molecular structure of 1,1,3,5,5-Pentaphenyl-1,3,5-trimethyltrisiloxane. It is the main constituent of the Dow Corning diffusion pump oils DC704 and DC705. Five phenyl and three methyl groups are regularly attached to the trisiloxane (\ce{Si3O2-R8}) backbone. The molecular dimensions can roughly be estimated using the diameter of a phenyl ring ($\sim 5.4$\,\AA) to be between 1\,nm and 1.5\,nm.\cite{Gruener2010}}
\label{DC705}
\end{figure}

As oily model liquids we employed the diffusion pump oils DC704 and DC705 from Dow Corning. Their main building blocks are siloxane molecules, see Fig. \ref{DC705} and exhibit a negligible vapor pressure and a relatively high viscosity (see Table~\ref{tab:propdcoils})\cite{Gruener2010}. 

\subsection{Experimental}
The capillary rise dynamics of the silicone oils have been recorded gravimetrically and by means of a CCD monochrome camera placed in front of the imbibition setup. The imbibition dynamics can be tracked by recording the increase in the sample's mass due to liquid uptake \cite{Gruener2010} \cite{Gruener2011} \cite{Gruener2016}. Such experiments can be easily performed by means of a laboratory scale. The sample is attached to the balance allowing time dependent recording of the gravitational force acting on the porous block -- see inset in Fig.~\ref{fig:imb_bsp} for {an} illustration. 

\begin{table*}[!t]
\centering
\begin{tabular}{c c c c}  
 DC   &  density $\rho$  &  viscosity $\eta$  &  surface tension $\sigma$    \\ 
    &   (g/ml) &   (mPa\,s) &   (mN/m)\\
   704 & 1.0772 &  48.99 &  32.85\\ 
  705 & 1.1033 &  218.8 &  35.24
\end{tabular}
\caption{{Fluid properties: {Dow Corning silicone oils} at $T=25\,$\dC. The values are taken from pycnometer and rheometer measurements as well as from Ref.~\cite{LB_IV_16}.}\label{tab:propdcoils}}

\end{table*}

\section{Darcy Analysis of Liquid Imbibition in a Hydrophilic Nanoporous Medium}
In the following we outline a phenomenological treatment of the capillary rise in a porous medium, as described in detail in Refs.~\cite{Gruener2016}\cite{Gruener2010}. It assumes that both gravitational and inertial affects are negligible compared to the huge capillary forces in the nanopores. 
However, the model considers changes in the available volume porosity for imbibition $\phi_{\rm i}$ compared to the nominal porosity of the sample $\phi_0$, as determined after bake-out in vacuum, since during the imbibition process pore space is occupied by preadsorbed water and hence no longer available. The exact value of the\rem{so-called} initial porosity $\phi_{\rm i}$ is accessible by an analysis of the overall mass uptake of the sample.

Moreover, the resulting changes in the curvature radius of the concave liquid menisci in the pores are considered as well as an effective hydraulic pore diameter $r_h$. This diameter does not necessarily have to agree with the pore radius $r_0$, determined by the sorption isotherm measurements of the baked-out sample, because of either strongly adsorbed, immobile boundary layers and thus a negative velocity slip length $b$ ($r_{\rm h}<r_0$, i.e. $r_h=r_0+b$)\cite{Gruener2011}, or due to velocity slippage at the pore walls ($r_{\rm h}>r_0$) \cite{Gruener2009}\cite{Gruener2009b}\cite{Vo2015}\cite{Vincent2016}\cite{Shen2018}. Only for the standard no-slip boundary condition is $r_{\rm h} \equiv r_0$. Note that $r_{\rm h}$ coincides with the radius over which a parabolic flow profile is established in the pore -- see Fig.~\ref{fig:VycorImbIllust}{(a)} 

Neglecting the initial ballistic imbibition regime \cite{Kornev2001}, which can be estimated to last just a few ps, we focus on the regime of viscous flow, where viscous dissipation prevails and acts against the capillarity-driven liquid uptake by the porous medium \cite{Bell1906}\cite{Lucas18}\cite{Washburn21}\cite{Rideal1922}. 
The competition of the constant Young-Laplace driving pressure and the increasing viscous drag in the liquid column behind the advancing imbibition front results in the Lucas-Washburn-Darcy $\sqrt{t}$ law for the rise height $h(t)$
\cite{Lucas18}\cite{Washburn21} \cite{Rideal1922} 
\begin{equation}
h(t) = {\sqrt{\frac{\sigma\, \cos\theta}{2\,\phi_{\rm i} \, \eta}} \; \Gamma} \; \sqrt{t} 
\label{eq:LWht}
\end{equation}
and thus also for the sample's mass increase $m(t)$ due to the liquid uptake $m(t)${,} 
\begin{equation}
m(t) = \underbrace{{ A  \;  \rho \; \sqrt{\frac{\phi_{\rm i} \,\sigma\, \cos\theta}{2\, \eta}} \; \Gamma}}_{C_{\rm m}} \; \sqrt{t} \; 
\label{eq:LWmt}
\end{equation}
where the proportionality constant becomes:
\begin{equation}
\Gamma = \frac{r_{\rm h}^2}{r_0} \;\sqrt{\frac{\phi_0}{\tau\,r_{\rm L}}}
\label{eq:Gamma}
\end{equation}
In the pre-factor $C_m$ of Eq. \ref{eq:LWmt} $\sigma$, $\eta$, $\rho$, $A$ refer to the surface tension, the shear viscosity, the density of the imbibing liquid, and the cross-sectional area $A$ of the sample in contact with the bulk liquid reservoir, respectively. The contact angle $\theta$ describes the wettability of the pore wall by the liquid. The tortuosity $\tau=3.6$ describes the connectivity and meandering of the pores in Vycor glasses \cite{Gruener2016}\cite{Lin1992}. Moreover, we introduce the quantity $\Gamma$ and term it Lucas-Washburn-Darcy imbibition ability. It is solely determined by matrix-specific quantities. Thus, it should be equal for matrices with identical internal structure and chemical composition -- independent of the imbibed liquid. It can be considered as a generalized pre-factor of the Lucas-Washburn law and thus of the imbibition speed. The larger $\Gamma$ the faster is the imbibition process. It is direct{ly} proportional to the square root of the pore dimensions expressed by a pore radius $r$ (as can be easily seen by applying $ r_{\rm h}= r_{\rm 0}= r_{\rm L}\equiv r $). Hence, the liquid will rise {faster in larger pores}.
 
The imbibition ability allows one to directly compare imbibition experiments with different liquids \cite{Gruener2016}\cite{Gruener2016b}. Furthermore, its absolute value contains information on the nanoscopic flow behaviour, especially on the hydrodynamic radius $r_{\rm h}$ and hence on the hydrodynamic boundary condition, which itself depends on the fluid/pore wall interaction. 

\section{Experimental Results and Discussion}
\begin{figure} \center
\includegraphics[angle=-90,origin=c, width=.8\columnwidth]{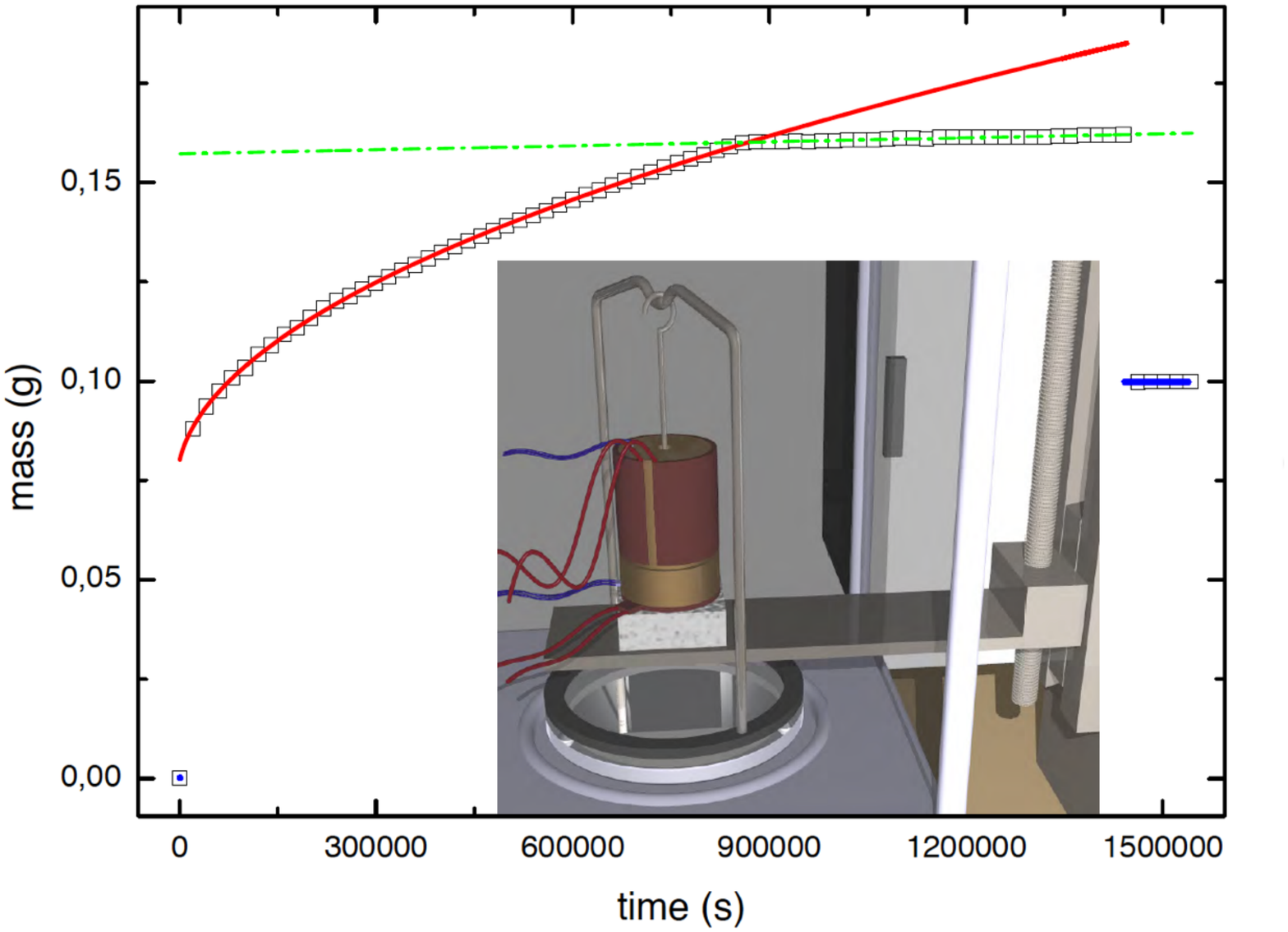}
\caption{Gravimetric Imbibition Experiment. Increase in mass (squares) of a porous Vycor block (V5) due to  imbibition of silicone oil (DC704) at room temperature. According to Eq.~\eqref{eq:LWmt} the prefactor of the\rem{shown} $\sqrt{t}$-Lucas-Washburn-Darcy kinetics fit (red solid line)\rem{comprises} {provides} information on the microscopic flow behavior expressed in terms of the imbibition ability $\Gamma$. The\rem{gradual} mass increase comes to a halt and a constant plateau (dashed line) is reached, when the sample is completely filled. For clarity {only every 1200th data point is shown}. Inset:\rem{Raytracing} Gravimetric imbibition setup\rem{employed} {used} for the capillary rise experiments.\cite{Gruener2010}\cite{Gruener2011}}
\label{fig:imb_bsp}
\end{figure}

The gravimetric experiments are illustrated by a representative mass increase measurement depicted in Fig.~\ref{fig:imb_bsp} for DC704 in V5. Four distinct regimes are observable. In the beginning the sample hangs freely above the bulk reservoir, $m(t)$=0. The measurement is started by moving the cell upward until the sample touches the liquid surface. A liquid meniscus forms at the outer perimeter of the sample rod, see picture in Fig.~\ref{fig:VycorImbIllust}(c). This induces a traction force F$_{S}$ acting on the porous matrix's surface towards the reservoir. For a given perimeter length $P$ of the meniscus it is determined by $F_S = P\,\sigma\,\cos{\theta_{0}}$, which for $\theta_0=0$\degr\ and the surface tension $\sigma \approx 33\,\frac{\rm mN}{\rm m}$ of silicone oil $T=20$\,deg C results in F$_S$ $\approx$ 0.7\,mN or, equivalently, in a mass jump of $\Delta m \approx 0.07\,$g, in good agreement with the measurement. 

This mass jump is a constant offset that does not affect the subsequent imbibition process. The latter is the outstanding effect in regime (b) and can also nicely be tracked by optical imaging, see the picture in Fig.~\ref{fig:VycorImbIllust} and the movie on the imbibition process in the supplemental. There is a difference in the optical refractive index between the oil filled and the empty porous glass. This results in changes in the light refraction at the advancing imbibition front. In combination with the cylindrical matrix shape this leads to an apparent macroscopic meniscus at the advancing oil front in an optical imaging experiment and thus allows one to track, in principal by bare eye, the imbibition front moving in the porous glass\cite{Gruener2010},\cite{Gruener2012}\cite{Gruener2016}. 

The mass uptake in regime (b) can be described by a $\sqrt{t}$-fit in accordance to Eq.~\eqref{eq:LWmt}, see Fig.~\ref{fig:imb_bsp}. Thus, the $\sqrt{t}$-fit provides via Eq.~\eqref{eq:LWmt} the imbibition ability. At some point a plateau is reached (regime (c)) indicating that the sample is completely filled with liquid. The blue line in regime (d) finally indicates the overall mass uptake of the liquid saturated sample.  
 
\begin{figure} \center
\includegraphics[angle=0,width=.8\columnwidth]{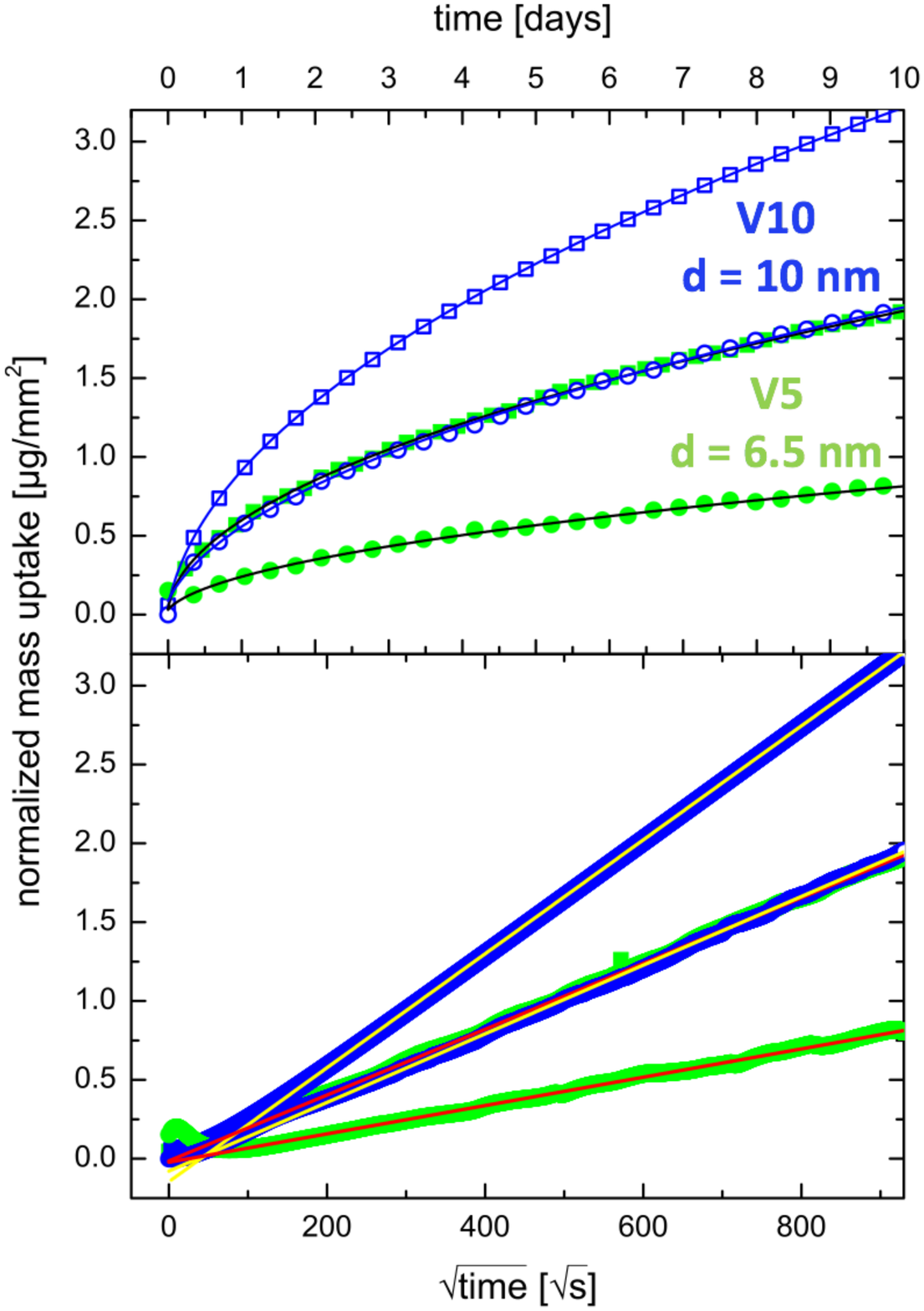}
\caption{Gravimetric experiment on silicone oil imbibition in monolithic nanoporous silica. (a) Plotted is the normalized mass uptake of Vycor glass V10 during imbibition of DC704 (open squares) and DC705 (open circles) as well as V5 upon DC704 (solid squares) and DC705 (solid circes) invasion, respectively. For clarity only every 6000th data point is shown. The solid lines correspond to $\sqrt{t}$-fits of the capillary rise dynamics as discussed in the text. (b) Normalized mass uptakes of panel (a) versus $\sqrt{t}$. The straight lines represent $\sqrt{t}$-fits from which the imbibition abilities $\Gamma$ are determined.}
\label{fig:imb_normalized}
\end{figure}

Analogous experiments are performed also for DC705 and V10 matrices, see Fig.~\ref{fig:imb_normalized} for the mass uptake of the corresponding liquid/matrix combination normalized by the cross-sectional area $A$ of the Vycor monolith ($A_{V5}$=36.96~mm$^2$, $A_{V10}$=25.23~mm$^2$ ). The solid lines represent $\sqrt{t}$-fits. Except for an initial phase of a few minutes, where presumably due to the formation and slow relaxation of the meniscus at the outer perimeter of the sample as well as buoyency effects due to the immersion in the viscous oils lead to deviations from the $\sqrt{t}$-scaling, the imbibition kinetics follows the Lucas-Washburn kinetics for several days. This evidences a remarkably robust description of the imbibition process by the L-W kinetics derived above. Except for the DC704/V10 and DC705/V5 systems which by coincidence almost agree in the imbibition dynamics, there are sizeable differences in the kinetics between the fluid/matrix combination resulting from the differing viscosities and mean hydraulic diameters. Note that in agreement with Eq.\ref{eq:Gamma} the liquid uptake rate for a given pore diameter increases with decreasing viscosity and for a given viscosity with increasing pore diameter.

The $\sqrt{t}$-fits of the mass-uptake curves, shown as solid lines in Fig.~\ref{fig:imb_normalized}, result according to Eq.~\ref{eq:LWmt} in the imbibition coefficients $C_m  =(7.6826\pm0.0009), (9.230\pm0.001), (3.2274\pm0.0016),  (5.2958\pm0.0007)\cdot 10^{-5}g/\sqrt{s}$. Along with the bulk fluid properties, see Tab.~\ref{tab:propdcoils}, the cross-sectional areas of the V5 and V10 samples (A$_{V5}$=36.96 mm$^2$,A$_{V10}$=25.23 mm$^2$ ) and the initial porosities $\Phi_i=0.275$ and 0.295 for V5 and V10, respectively, these experimentally determined values allow then to determine the corresponding imbibition abilities, \textit{i.e.} according to Eq.~\ref{eq:LWmt} ${\Gamma}= C_m/( A  \;  \rho \; \sqrt{\frac{\phi_{\rm i} \,\sigma\, \cos\theta}{2\, \eta}})$. From these imbibition abilities the effective hydraulic radius $r_h$ is determined with Eq.~\ref{eq:LWmt} and therefrom the slip lengths $b=r_h-r_0$, see Tab.~\ref{tab:GammaM_DCoils} for the resulting imbibition abilities and slip lengths. 
 
These slip lengths are all negative, indicating a reduced hydraulic radius $r_h$ and thus indicate a sticking boundary layer, whose thickness is approximately 1.4\,nm. The mean value determined from the 4 experiments is $\bar{b}=-(1.4\pm0.3$)~nm. Interestingly this result is independent of both the sample type and the silicone oil. This encourages us to attribute this observation, similarly as in previous studies on water, hydrocarbon, and liquid crystal imbibition in silica \cite{Gruener2009}, \cite{Gruener2009a}, \cite{Gruener2011}, \cite{Gruener2016}, \cite{Gruener2016b} to the formation of an immobile layer of strongly adsorbed molecules, that substantially lowers the invasion dynamics. 

Arguably, the largest uncertainty in the determination of the slip length results from the ambiguities in the concept and the measurement of the tortuosity \cite{Levitz1998}. To date several techniques have been applied to extract the tortuosity of the isotropic pore network in Vycor glass. Deducing the diffusion coefficient of hexane and decane by means of small angle neutron scattering (SANS) measurements $\tau$ was found to be in the range of 3.4 - 4.2 \cite{Lin1992}. Calculations based on three-dimensional geometrical models for Vycor yielded a value of approximately 3.5 \cite{Crossley1991}\cite{Levitz1998}. We chose a value of $\tau$=3.6 with an uncertainty of $\pm$0.5. A tortuosity of about three seems reasonable if one considers that in an isotropic medium such as Vycor, the porosity can, in first approximation, be accounted for by three sets of parallel capillaries in the three spatial directions; but only one third of these capillaries sustain the flow along the pressure drop. Hence the hydraulic permeability in the Darcy sense is reduced by a factor of 3. A value larger than three reflects the extended length of a meandering capillary beyond that of a straight one. According to Eq.\ref{eq:Gamma} $\tau$ scales $\propto (r_0+b)^4$ for a fixed $\Gamma$ and mean radius. Thus, large uncertainties in $\tau$ result in comparably small changes in the slip lengths determined in our experiments, and, vice versa, small changes in the slip length $b$ necessitate significantly altered $\tau$s. For example, to explain the measured $\Gamma$s in our experiment with the standard no-slip boundary condition $b=0$ (no sticky boundary layer), we would need $\tau=$ 27, 38, 13, and 11 to explain the $\Gamma$s for the V5/DC704, V5/DC705, V10/DC704 and V10/DC705 fluid/matrix combinations, respectively. These values are unreasonable. Moreover, we would need for the identical sample type (V5 or V10) values which differ significantly as a function of employed silicone oil. By contrast, the consistent use of $\tau$=3.6$\pm$0.5 for the analysis of all 4 fluid/pore-radius combinations results in quite similar slip lengths, i.e. $\bar{b}=-(1.4\pm0.3$)~nm.

Note that Cai and Yu extended the classic Lucas-Washburn law towards the consideration of flow heterogeneity in porous media originating in the tortuosity of a porous medium \cite{Cai2011}. This results in a modified time scaling of the invasion kinetics, $h(t) \propto t^{\frac{1}{2 D_{\rm T}}}$, where $D_{\rm T}$ is the fractal dimension of tortuosity in the Cai-Yu considerations with $1 < D_{\rm T} < 3$. However, except for deviations in the initial imbibition times for the most viscous oil, which we trace to experimental artifacts related to the approach of the bulk reservoir, we observe a robust Lucas-Washburn kinetics, i.e. $D_{\rm T}=1$ for all fluid/media combinations studied here. Therefore, we conclude that in our experiments tortuosity-induced flow heterogeneities are negligible and the classic Lucas-Washburn law ($D_{\rm T}=1$) is valid.  

\begin{table*}[!t]
\centering
\begin{tabular}{C{\linewidth/6}*{9}{C{\linewidth/6}}}
   Liquid   & Vycor V5 & & Vycor V10 \\
       \\
       & ${\Gamma}$ ($10^{-7}\,\sqrt{\rm m}$)  & $b$  (nm)              & ${\Gamma}$($10^{-7}\,\sqrt{\rm m}$) & $b$ (nm)	\\				         DC704  & $63.6\pm 5.6$       &  $-1.35\pm 0.23$   & $109.2\pm 12.6$    & $-1.39\pm 0.40$  \\
  DC705  & $53.2\pm 4.8$       &  $-1.52\pm 0.21$   & $122.4\pm 11.4$    & $-1.19\pm 0.38$  \\ 
\end{tabular}
\caption{Characterization of the imbibition dynamics of Dow Corning silicone oils in porous Vycor at room temperature by means of the imbibition ability ${\Gamma}$ and the resultant slip length $b$.}
\label{tab:GammaM_DCoils}
\end{table*}

At first glance, the oily character of the fluids employed may suggest a polymeric rheology, and thus one may have expected faster flow kinetics than predicted by a no-slip velocity boundary condition \cite{Priezjev2004} \cite{Dimitrov2007} \cite{Hatzikiriakos2012}, and thus positive slip lenghts. However, the molecular structure of the siloxanes studied here is not really flexible, chain-like but rather rigid, so that the observation of a negative slip length is not too surprising. Moreover, in agreement with previous studies on simple wall-wetting liquids and thus strong attractive liquid-solid interaction sticky layers and thus negative $b$s, whereas for non-wetting situations rather positive and thus positive $b$s are expected \cite{Vinogradova99}\cite{Lasne08}\cite{Barrat99}\cite{Baeumchen2012}\cite{Wu2017}.  

From previous analogous experiments on hydrophilic Vycor glasses \cite{Gruener2009}\cite{Gruener2016} we suggest that independent of the respective silicone oil a monolayer of water directly adjacent to the pore walls is most likely an essential part of this sticking layer. The water coating of the pore walls is due to the\rem{final} {finite} humidity in our experiments of about 30$\%$. It results in the formation of water layers adsorbed at the pore wall. Especially the first adsorbed water layer is stabilized by the attractive potential between the hydroxylated silica and the polar water molecules. This is also indicated by a pronounced monolayer step in sorption isotherms and a slow self-diffusion dynamics \cite{Bonnaud2010} of water in hydroxylated nanopores. The sticky water layer is highly stabilized and cannot be displaced by the silicone molecules. It corresponds to a thickness of $\sim$ 0.25 nm \cite{Gruener2009}. The remaining part of the sticking layer thickness $b$ can then be attributed to a second pinned layer composed of molecules of the silicone oils. Here, the block-like molecules are presumably arranged parallel to the pore walls (flat-lying), which results in an overall immobile boundary layer thickness of $\sim 1.4$~nm$=(0.25+1.15)$~nm, in good agreement with the negative $b$s observed. 

For the mobile, inner region (away from the interface) classical concepts of hydrodynamics based on continuum-like properties such as shear viscosity and surface tension remain valid and we also find no hints for shear thinning or thickening. It would result in deviations from the $\sqrt{t}$-L-W dynamics \cite{Cao2016}, because of the variation of the shear rate during the imbibition process. 

The partitioning in immobile interfacial liquid layers and bulk-like fluidity in the pore centers are also in accord with measurements on the self-diffusion dynamics of other, simple liquids in nanopores. \cite{Huber2015} For example, one component with\rem{a} bulk-like self-diffusion dynamics and a second one which is immobile, {and thus `}sticky{',} has been found in quasi-elastic neutron scattering experiments \cite{Kusmin2010}\cite{Kusmin2010a}\cite{Hofmann2012}. 

The formation of layered structures of silicone oils at capillary walls has also been inferred from imbibition experiments performed for silicone oil in macroscopic borosilicate capillaries \cite{Wu2017b}. As discussed in detail in Ref.~\cite{Wu2017b} it can result in sizeable changes in the contact angle as a function of the advancing menisci and thus to reduced imbibition velocities. Note, however, that here the meniscus velocities are orders of magnitude smaller than in macroscopic capillaries. Therefore, we assume unchanged contact angles close to 0 and trace the reduced imbibition velocities solely to the formation of an immobile boundary layer.

It is interesting to remark that a significant light scattering is observable at the invasion front, in particular in the late stages of the imbibition process after several days, see the imbibition video in the supplementary. As is outlined in detail in Ref.~\cite{Gruener2016} it can be traced to the disorder of the Vycor pore structure, in particular the pore size distribution. The resulting Laplace pressure and hydraulic permeability variations result in the formation of filled and still empty pore space volumes,\cite{Clotet2016} whose characteristic extensions are in the order of visible light wavelengths. Along with the difference in the refractive index between filled and empty pores, this leads to light scattering and thus in a whitening of the imbibition front. Neutron imaging experiments \cite{Gruener2012}\cite{Gruener2016} in accord with pore network simulations and a scaling theory for the long time behavior of spontaneous imbibition in porous media consisting of interconnected pores with a large length-to-width ratio, as in Vycor, could trace the front broadening to a complex dynamics of the individual menisci. In particular long-lasting meniscus arrests, when at pore junctions the meniscus propagation in one or more branches comes to a halt when the Laplace pressure of the meniscus exceeds the hydrostatic pressure within the junction. Unfortunately, such single pore invasion events are not directly visualisable in nanoporous media so far, despite the substantial improvements in 3D X-ray tomography to resolve such events \cite{Berg2013}. However, microfluidic experiments corroborate the scenario \cite{Sadjadi2015} outlined above. 

Note, that the mean position of the imbibition front in Vycor is still well defined \cite{Gruener2012}\cite{Gruener2016}, despite the sizeable capillarity and permeability disorder, see Ref.~\cite{Gruener2016} for a detailed analysis of the imbibition front broadening for V5 and V10. Therefore, the measurement of the time-dependent gravimetric mass uptake, performed here, is a robust method to quantify the invasion kinetics in the complex, 3D interconnected pore network of Vycor glasses.

\section{Conclusions}
We experimentally explored spontaneous imbibition of silicone oils in mesoporous, monolithic silica glass. The kinetics follow Lucas-Washburn laws typical of spontaneous imbibition in macroporous media with homogeneous porosity. A Darcy analysis indicates that imbibition speeds can be quantitatively traced to bulk fluid parameters, if {a} {sticking} boundary layer of 1.4~nm \rem{corresponding to the thickness of a} of flat-lying silicone molecules \rem{backbone} and a monolayer of water adsorbed at the pore walls is assumed. These rheological insights are in good agreement with previous experiments on simple, wetting liquids in nanoporous media  \cite{Huber2007}\cite{Gruener2009}\cite{Gruener2009a}\cite{Gruener2011}\cite{Gruener2016}\cite{Vincent2016}\cite{Koklu2017}. They are also corroborated by simulation studies on liquid transport in nanopores \cite{Dimitrov2007} \cite{Vo2015}\cite{Vo2016}\cite{Zhang2018b} and phenomenological models for spontaneous imbibition in nanoporous media. \cite{Shen2018}\cite{Feng2018}

Future studies on the polarity and hydration state dependent flow across these silica monoliths would also be particularly interesting. These surface characteristics should have sizeable effects on the hydrodynamic boundary condition and eventually lead to a mobilization of the sticky boundary layer \cite{Gruener2016}\cite{Sendner2009}. Also complementary measurements on the self-diffusion of the oil molecules, e.g. by quasi-elastic neutron scattering \cite{Kusmin2010}\cite{Kusmin2010a} or nuclear magnetic resonance, \cite{Valiullin2006a} could give important information on the stochastic molecular motions and via the Stokes-Einstein relation on the flow viscosity in the spatially-confined geometries.

The complex sponge-like geometry of our nanoporous media resemble on the one hand many technological and natural porous systems, on the other hand experiments on straight independent nanopores would reduce experimental ambiguities with regard to pore size distributions and tortuosity \cite{Elizalde2014} and are thus desirable. 

The pore space in tight oil reservoirs, in particular shale, can often roughly be segregated in inorganic hydrophilic pores and organic hydrophobic pores \cite{Yassin2017}\cite{Zhang2018a}. Even for macroscopic porosity, this dual-wettability results in particular complex front movements, where for example local instabilities control the progression of invading interfaces \cite{Singh2017}. In nanoporous systems the wettability-induced variation in hydrodynamic boundary conditions (negative and positive velocity slippage) could therefore induce even more complex front movement dynamics.

Also multiscale porosity is present in many oil reservoirs, soils, rocks, and shales and materials such as concrete with a pore distribution spanning orders of magnitude, frequently from the macroscale down to sub-nm dimensions. Therefore, it would be also particularly interesting to explore spontaneous imbibition in sub-nm pores in the future. There, a non-Darcy behavior is expected \cite{Falk2015}, because of a complex interplay of adsorption, stochastic motions and viscous flow. Moreover, the advent of materials with well-defined hierarchical porosity \cite{Hartmann2016} may offer the possibility to explore multiscale descriptions of this complex interplay of transport mechanisms.\cite{Botan2015}


\begin{acknowledgements}
This work has been supported by the Deutsche Forschungsgemeinschaft (DFG), project Hu850/9-1, ''Oxidic 3d scaffold structures for wetting-assisted shaping and bonding of polymers''.
\end{acknowledgements}

\bibliographystyle{spphys}       



\end{document}